# Entre choix et contraintes : décryptage de l'empouvoirement financier chez les femmes entrepreneures en France

## Jonathan Labbé

*Jonathan Labbé est maître de conférences à l'IAE de la Nancy School of Management. Ses recherches, menées au sein du laboratoire CEREFIGE, se concentrent sur la finance entrepreneuriale, avec un intérêt particulier pour les relations entre capital-investisseurs publics et privés.*

IAE, Nancy School of Management
Laboratoire CEREFIGE (EA3942)
86, rue Sergent-Blandan
54000 NANCY, France
**jonathan.labbe@univ-lorraine.fr**

## Typhaine Lebègue

*Typhaine Lebègue est maîtresse de conférences en sciences de gestion à l'IAE Tours Val de Loire. Ses recherches portent principalement sur l'entrepreneuriat des femmes. Elle est cofondatrice d'un groupe de recherche sur l'entrepreneuriat et le genre.*

IAE Tours Val de Loire
Laboratoire VALLOREM (EA 6296)
50, avenue Jean-Portalais
BP 0607
37206 TOURS CEDEX 3, France
**typhaine.lebegue@univ-tours.fr**

## Abdel Malik Ola

*Abdel Malik Ola est maître de conférences en sciences de gestion à l'IAE Tours Val de Loire. Ses recherches portent sur la compréhension de la décision d'investissement, avec un intérêt pour les dimensions psychocognitives de la décision.*

IAE Tours Val de Loire
Laboratoire VALLOREM (EA 6296)
50, avenue Jean-Portalais
BP 0607
37206 TOURS CEDEX 3, France
**abdelmalik.ola@univ-tours.fr**

## Résumé

*Cette recherche examine l'empouvoirement des entrepreneures dans le cadre du financement entrepreneurial en France. Elle s'interroge sur les facteurs permettant à certaines entrepreneures d'accéder plus facilement à des catégories spécifiques de financement externe. Le cadre théorique mobilisé repose sur le concept d'empouvoirement, exploré à travers ses dimensions personnelle et relationnelle. L'étude s'appuie sur une approche quantitative,*

*utilisant des données issues d'une enquête représentative menée auprès de dirigeantes d'entreprise. Les résultats montrent que le statut de dirigeante fondatrice influence différemment l'accès aux financements externes : il augmente les chances de réussir une levée de fonds, mais réduit celles d'obtenir un financement bancaire. Ce constat souligne l'importance des dynamiques d'empouvoirement, qui varient selon le type de financement. En outre, des caractéristiques telles que la présence du conjoint dans l'entreprise, un revenu élevé, l'appartenance à un réseau professionnel et la mixité de ce réseau complètent l'analyse des disparités d'accès. Cette première étude en France sur ce sujet propose des pistes pour enrichir la compréhension de la diversité des situations vécues par les dirigeantes fondatrices contribuant ainsi à déconstruire l'image homogène de l'entrepreneuriat des femmes.*

**Mots-clés**

**Femmes entrepreneures, Empouvoirement, Financement externe, Financement bancaire, Levée de fonds**

**Choices or constraints: decoding financial empowerment among women entrepreneurs in France.**

**Abstract**

*This research examines the empowerment of women entrepreneurs in the context of entrepreneurial financing in France. It explores the factors that allow some women entrepreneurs to access certain categories of external finance more easily. The theoretical framework used is based on the concept of empowerment, explored through its personal and relational dimensions. The study relies on a quantitative approach, using data from a representative of women entrepreneurs. The results show that the status of a founder affects access to external finance in different ways: it increases the chances of successful fundraising, but reduces the chances of obtaining bank finance. This finding highlights the importance of empowerment dynamics, which vary according to the type of financing. In addition, characteristics such as the presence of a spouse in the business, high income, membership of a professional network and the diversity of this network complete the analysis of inequalities in access. This study, the first of its kind in France, suggests ways of enriching our understanding of the diversity of situations experienced by female founders, thus helping to deconstruct the homogeneous image of women's entrepreneurship.*

# INTRODUCTION

Si les femmes ont été longtemps largement sous-représentées dans la part des entrepreneurs, elles occupent une place croissante (Brush, Greene, Balachandra et Davis, 2018 ; Poczter et Shapsis, 2018) et dirigent des entreprises qui affichent des performances égales ou meilleures à celles des hommes (Aernoudt et De San José, 2020 ; Villaseca, Navío-Marco et Gimeno, 2020). Toutefois, la littérature met en avant des phénomènes de discrimination envers les entrepreneures dans l'accès aux financements externes, pourtant cruciaux à la croissance des entreprises (Basiglio, De Vincentiis, Isaia et Rossi, 2023 ; Wesemann et Wincent, 2021). Cet accès difficile au financement s'explique notamment par le fait que les entrepreneures sont présumées manquer de confiance en elles, avoir une forte aversion au risque, posséder de faibles compétences managériales et de négociation et appartenir à un réseau professionnel faiblement connecté à l'écosystème des investisseurs (Comeig, Holt et Jaramillo-Gutiérrez, 2022 ; Kwapisz et Hechavarría, 2018).

Les causes énoncées sont liées à l'influence des stéréotypes de genre sur la faculté des financeurs à interpréter les projets des femmes (Balachandra, Briggs, Eddleston et Brush, 2019 ; Eagly et Karau, 2002 ; Liao et al., 2024). Les stéréotypes de genre sont des croyances partagées sur les attributs et comportements typiquement associés aux hommes et aux femmes, incluant des traits de personnalité ainsi que des attentes en matière de comportement (Eagly et Karau, 2002). Ainsi, les femmes qui occupent des rôles masculins comme les entrepreneurs sont perçues comme violant les rôles de genre (Liao et al., 2024), ne possédant pas les caractéristiques masculines traditionnellement valorisées pour susciter l'intérêt et le soutien de la part des fournisseurs de ressources financières (Balachandra et al., 2019 ; Liao et al., 2024). Leur capacité à entreprendre peut être questionnée par les professionnels du financement, qui les perçoivent comme moins compétentes (Alsos et Ljunggren, 2017 ; Malmström, Voitkane, Johansson et Wincent, 2020) ou craignent un jugement négatif d'approbation sociale de la part de leurs pairs investisseurs (Liao et al., 2024). Brush et al. (2018) concluent que les femmes manquent à la fois de préparation et de motivation pour attirer les investisseurs en capital risque. Les femmes finissent par intérioriser cette perception sociale en leur défaveur ; elles s'estiment moins légitimes que leurs homologues masculins et s'auto-limitent dans leur recherche de financements externes (Brush et al., 2018 ; Carter, Brush, Greene, Gatewood et Hart, 2003 ; Dutta et Mallick, 2023).

Cette comparaison permanente avec leurs homologues masculins ne permet pas l'identification des véritables éléments influençant l'obtention de financements par les entrepreneures. Les femmes entrepreneures ne devraient pas « exister » que par rapport aux hommes, mais elles devraient plutôt être considérées comme des unités d'analyse indépendantes. Elles ont un parcours entrepreneurial et font des choix qui incarnent les nouvelles valeurs qu'elles souhaitent porter (Brière, Auclair et Tremblay, 2017 ; Digan, Sahl, Mantok et Patel, 2019). L'action d'entreprendre peut déclencher un processus d'empouvoirement à l'issue duquel les femmes deviennent suffisamment conscientes de leurs compétences pour pouvoir les défendre (Cattaneo et Chapman, 2010 ; Desai, Chen, Reddy et McLaughlin, 2022 ; Malmström, Burkhard, Sirén, Shepherd et Wincent, 2023). Ce processus d'empouvoirement représente un cheminement par lequel « celles qui ont été privées de la capacité de faire des choix de vie stratégiques acquièrent cette capacité » (Kabeer, 1999). Selon Huis, Hansen, Lensink et Otten (2020), le processus d'empouvoirement intègre un niveau personnel qui renvoie à une meilleure perception de soi, mais également un niveau relationnel à travers l'encastrement social des femmes. En outre, l'empouvoirement varie d'une femme à l'autre (Ahl, 2006 ; Henry, 2021). Les recherches soulignent qu'il est crucial de prendre en compte les divers aspects de l'identité de la femme entrepreneure et son intégration dans de multiples groupes sociaux afin de comprendre l'unicité de leurs expériences (Martinez Dy, 2021 ; Nelson et Constantinidis, 2017). Aussi, les femmes entrepreneures présentent des différences, que ce soit au niveau des caractéristiques de l'entreprise ou de leurs propres traits (Constantinidis, Cornet et Asandei, 2006), ce qui influence probablement aussi leur approche en matière de financement entrepreneurial (Brush et al., 2018 ; Cowling, Marlow et Liu, 2020 ; Malmström, Johansson et Wincent, 2017).

Dans cette recherche, nous rompons avec la comparaison traditionnelle entre les hommes et les femmes pour nous inscrire dans la perspective de l'empouvoirement des femmes entrepreneures. Nous cherchons à expliquer pourquoi certaines femmes entrepreneures, en l'occurrence les femmes dirigeantes fondatrices, ont davantage accès à certains types de financements externes. Le choix de se focaliser sur les femmes dirigeantes fondatrices s'explique par le fait qu'elles vont expérimenter ce processus d'empouvoirement en raison de leur rôle actif dans la création et la gestion courante de leur entreprise (Boumedjaoud et Messeghem, 2020 ; Richomme-Huet et d'Andria, 2012). À notre connaissance, cette étude constitue la première analyse quantitative de type en France depuis les travaux d'Orhan (2001) sur les discriminations vécues par les entrepreneures françaises dans l'accès au crédit bancaire. Elle approfondit les connaissances établies en considérant les deux niveaux d'empouvoirement, personnel et relationnel (Digan et al., 2019 ; Desai et al., 2022 ; Huis et al., 2020). Alors que le niveau relationnel est mesuré à travers les caractéristiques du réseau professionnel, le niveau personnel est appréhendé à travers l'engagement opérationnel et financier des entrepreneurs. Nous montrons comment les deux niveaux d'empouvoirement influencent différemment les probabilités d'accès au financement externe.

L'article se structure autour de trois sections. La première section présente notre cadre de recherche. En mobilisant la littérature sur les types de financements externes retenus, nous mettons en lumière la présence d'ambiguïtés concernant l'influence du genre. Nous mobilisons ensuite le concept d'empouvoirement pour dévoiler le processus par lequel les femmes entrepreneures développent leur capacité d'action vis-à-vis de l'accès au financement. L'empouvoirement invite à envisager la prise en compte des niveaux personnel et relationnel dans l'identification des aptitudes développées par les entrepreneurs. La deuxième section décrit notre méthodologie en détaillant l'échantillon, les mesures et les méthodes utilisées pour l'analyse des données. Enfin, la troisième section présente et discute les résultats et leurs implications.

## 1. REVUE DE LITTÉRATURE

### 1.1. Des résultats ambigus selon les différents canaux de financement

La littérature sur le financement des entrepreneures s'intéresse plus au crédit bancaire qu'aux levées de fonds auprès d'investisseurs en capital. Si la banque est le premier partenaire financier des entrepreneures (Eddleston, Ladge, Mitteness et Balachandra, 2016), le lien entre le genre et le rationnement du crédit bancaire n'a pas été clairement défini (Liu et Cowling, 2023 ; Malmström *et al.*, 2023 ; Qi, Ongena et Cheng, 2022). Alors que Wellalage et Thrikawala (2021) montrent que les femmes demandent et obtiennent moins de crédit bancaire que les hommes, d'autres auteurs réfutent l'existence d'un effet genre sur l'accès à cette catégorie de capitaux, observant qu'elles en font simplement moins fréquemment la demande (Basiglio *et al.*, 2023 ; Cornet et Constantinidis, 2004 ; Galli, Mascia et Rossi, 2020). Cornet et Constantinidis (2004) révèlent que plus de la moitié des femmes n'ont déposé aucune demande de financement bancaire durant les cinq premières années de vie de leur entreprise. Campanella et Serino (2019) montrent qu'elles en font bien la demande, mais reçoivent un nombre plus élevé de refus de la part des banques. Par ailleurs, les pratiques bancaires ne semblent pas être liées au genre ni au Canada (Madill, Riding et Haines, 2006), ni au Royaume-Uni (Cowling, Marlow et Liu, 2020), ni aux États-Unis (Eddleston *et al.*, 2016), ni en Belgique (Cornet et Constantinidis, 2004). Pour autant, Bellucci, Borisov et Zazzaro (2010), Mascia et Rossi (2017) ainsi qu'Orhan (2001) affirment qu'une discrimination existe du fait que les femmes expérimentent des conditions d'accès au crédit bancaire plus défavorables. En examinant le niveau de confiance, Liu et Cowling (2023) concluent que tant l'excès que le manque de confiance sont préjudiciables aux femmes par rapport aux hommes ; un niveau modéré de confiance en leurs capacités peut en revanche avoir un effet positif sur leur probabilité d'accès au financement bancaire.

Du côté de la levée de fonds, le genre est désormais proposé comme un facteur déterminant dans le financement auprès d'investisseurs (Brush et al., 2018 ; Greene, Brush, Hart et Saparito, 2001 ; Liao et al., 2024). De façon générale, un faible nombre d'entrepreneures ont recours aux levées de fonds (Alsos et Ljunggren, 2017). Cependant, lorsqu'elles en font la demande, les femmes lèvent plus de fonds, quelles que soient les étapes du développement des projets (Aernoudt et De San José, 2020). Par ailleurs, elles sous-valorisent leur projet, cherchent à lever moins de capitaux auprès des business angels et des capital-risqueurs et offrent en contrepartie une part plus importante de leur société (Kwapisz et Hechavarría, 2018 ; Poczter et Shapsis, 2018). Chez les professionnels de l'investissement, il existerait une résistance cognitive lors de l'évaluation des projets des femmes, dévalorisant ces dernières de même que leur initiative entrepreneuriale (Harrison et Mason, 2007 ; Liao et al., 2024). L'esprit d'initiative et l'ambition, ainsi que la légitimité des femmes, sont remis en question tandis que les candidatures masculines sont appréciées sous un angle plus valorisant (Edelman, Donnelly, Manolova et Brush, 2018). Les projets des femmes entrepreneures reçoivent des évaluations inférieures à celles des hommes parce que leurs compétences sont sous-estimées par les investisseurs en capital (Amatucci et Sohl, 2004 ; Snellman et Solal, 2023).

Les questions autour de la capacité des femmes entrepreneures nous conduisent à examiner les possibilités pour elles d'accéder à différents types de financement externe, en utilisant le cadre de l'empouvoirement.

## 1.2. L'empouvoirement des femmes entrepreneures

De nombreux travaux sur la relation entre entrepreneuriat et genre ont posé la question en termes de pouvoir, souvent pour analyser l'entrepreneuriat comme moyen, lieu et pratique de domination. Ces recherches défendent l'idée que l'entrepreneuriat peut constituer un espace de résistance et d'empouvoirement pour les femmes. L'empouvoirement renvoie au processus par lequel les femmes développent leur capacité d'action et acquièrent du pouvoir (Kabeer, 1999) leur permettant d'améliorer leur vie individuelle, leur vie communautaire et contribuant à la société dans son ensemble (Cattaneo et Chapman, 2010). Nous revenons ensuite sur les spécificités du processus d'empouvoirement des entrepreneures avant d'aborder les dimensions que nous mobilisons.

### *1.2.1. L'empouvoirement des entrepreneures*

Chez les femmes, l'expérience entrepreneuriale permet d'atteindre un meilleur niveau de confiance en soi et de capacité psychologique (Brière, Auclair et Tremblay, 2017 ; Digan *et al.*, 2019). Au fil du temps, elles développent davantage d'empouvoirement, en prenant confiance en leur identité entrepreneuriale et en leur force de décision. Elles font croître leur capacité à accéder aux ressources et à les contrôler (Kivalya et Caballero-Montes, 2023). Les femmes dirigeantes ont souvent dû faire preuve de résilience pour s'imposer dans un environnement entrepreneurial dominé par des stéréotypes masculins (Kelly et McAdam, 2023). Il y a selon les termes de Barragan, Erogul et Essers (2018), une microémancipation par l'action d'éloignement vis-à-vis d'un statut que le monde extérieur leur adosse. Subjectivement, elles peuvent adopter des comportements entrepreneuriaux acceptés par les autres acteurs de l'entrepreneuriat. Aussi, les femmes dirigeantes, parce qu'elles ont une expérience entrepreneuriale leur permettant notamment d'accumuler des ressources qui favorisent la prise de pouvoir et la capacité à faire des choix (McAdam, Harrison et Leitch, 2019), vont avoir une autoperception élevée de leur empouvoirement. Il ressort de ces développements un accroissement des capacités décisionnelles, de choix et une évolution de la perception de soi dans le processus d'empouvoirement.

Nous pouvons donc considérer l'empouvoirement non pas comme une conséquence de l'entrepreneuriat, mais plutôt comme un point de départ des comportements entrepreneuriaux futurs. Par là, notre travail s'inspire de la perspective de Digan *et al.* (2019) qui, à notre connaissance, est l'une des rares à s'ancrer dans cette approche. Le comportement entrepreneurial qui nous intéresse est l'accès au financement externe. Nous estimons qu'en jouant sur les niveaux personnel et relationnel, l'empouvoirement est susceptible d'influencer les probabilités d'accès à différentes sources de financement externe.

### 1.2.2. Les dimensions de l'empouvoirement des femmes

L'empouvoirement revêt un caractère multidimensionnel, à la fois individuel et collectif (Desai *et al.*, 2022). Le processus d'empouvoirement des femmes peut être classé en trois catégories distinctes, mais interconnectées (Huis, Hansen, Otten et Lensink, 2017 ; Huis *et al.*, 2020) :

1. au **niveau personnel** se définit comme l'appréciation par les femmes de leurs capacités et de leurs compétences, ce qui se traduit par une plus grande estime de soi, une plus grande confiance en soi et une perception plus forte de sa légitimité (Digan *et al.*, 2019 ; Huis *et al.*, 2020 ; Kato et Kratzer, 2013). Ces éléments peuvent être considérés comme un point de départ important avant que les individus puissent acquérir le pouvoir d'influencer leur propre vie (Rowlands, 1998) ;
2. au **niveau relationnel** fait référence à l'idée que le pouvoir d'un individu ne peut être compris qu'en relation avec un autre individu ou un groupe (Anderson, John et Keltner, 2012 ; Thibaut et Kelley, 1959). Ainsi, le pouvoir ou la capacité d'un individu à influencer les autres dépend de ses relations spécifiques (Anderson, John et Keltner, 2012). Les recherches montrent que les relations proches, dans lesquelles les femmes sont intégrées, sont déterminantes pour leur empouvoirement (Belcher, Peckuonis et Deforge, 2011 ; Huis *et al.*, 2020). Concernant les entrepreneures, elles sont intégrées dans un réseau de relations appartenant à différentes sphères, notamment leur famille, leur communauté et leur partenariat professionnel, ce qui peut influencer différemment leur comportement entrepreneurial (Brush, de Bruin et Welter, 2009 ; Lebègue, 2015). Dans leur étude, Huis *et al.* (2020) concluent que la relation étroite avec le conjoint est prépondérante dans la capacité des femmes à exprimer des signes d'empouvoirement en termes de décisions financières ;
3. au **niveau sociétal** renvoie à la situation des femmes dans la société, comme leur représentation au Parlement ou dans la création d'entreprise. Cette dernière forme d'empouvoirement interagit avec les deux précédentes (Ng, Wood et Bastian, 2022).

Dans notre recherche, nous nous concentrons sur les deux premières formes et postulons que le comportement des entrepreneures vis-à-vis du monde extérieur, notamment les financeurs, renvoie à un processus dans lequel l'empouvoirement joue à la fois au niveau personnel et relationnel.

## 1.3. L'empouvoirement, engagement et financement

### *1.3.1. Empouvoirement et engagement entrepreneurial*

L'engagement perçu chez un entrepreneur, qui renvoie à la passion et l'énergie investies dans les comportements entrepreneuriaux, constitue un indicateur du degré de sa préparation vis-à-vis de son projet, de sa motivation et de sa capacité à surmonter les obstacles à venir (Cardon, Grégoire, Stevens et Patel, 2013 ; Chen, Yao et Kotha, 2009). L'engagement va influencer la façon dont le porteur de projet présente sa vision, sa confiance en sa capacité à réussir, ainsi que le degré d'implication des parties prenantes nécessaires au développement de son entreprise (Taylor, 2019).

Les femmes dirigeantes développent un capital psychologique qui démontre leur aptitude à accéder aux ressources et à les combiner pour atteindre des objectifs de performance (Digan *et al.*, 2019). L'engagement entrepreneurial des femmes implique la capacité à se faire accepter comme *leader* dans une équipe et à être capable de la diriger avec une bonne écoute (Richomme-Huet et d'Andria, 2012). Le processus d'empouvoirement va permettre aux entrepreneures à succès d'avoir suffisamment confiance en leur aptitude à décider, à faire des efforts et engager les ressources nécessaires pour atteindre les objectifs fixés (Digan *et al.*, 2019). Ainsi, d'une part, le niveau d'empouvoirement suppose qu'elles auraient les connaissances cognitives et le capital psychologique nécessaires pour l'entrepreneuriat. D'autre part, l'engagement démontre qu'elles peuvent mobiliser leurs connaissances, leurs compétences et leurs ressources cognitives pour atteindre l'objectif du projet. Par conséquent, il est possible de déduire que le processus d'empouvoirement peut permettre aux entrepreneures d'afficher un plus fort engagement vis-à-vis de leur projet.

### *1.3.2. Engagement et accès au financement externe*

L'engagement d'un entrepreneur vis-à-vis de son projet est un facteur important dans le processus d'accès au financement (Dutta et Mallick, 2023 ; Eddleston *et al.*, 2016). Il est souvent étudié dans les levées de fonds parce que le modèle de l'investissement privé se fonde sur la qualité de la relation avec les dirigeants postfinancement (Maxwell et Lévesque, 2014 ; Taylor, 2019). L'engagement est un indicateur du capital humain mobilisable dans le management des projets (Greene *et al.*, 2001). Son influence est définie de deux manières contradictoires.

Les femmes dirigeantes, qui vont démontrer de l'engagement entrepreneurial, sont dévalorisées parce qu'elles ne seraient pas en cohérence avec le stéréotype socialement reconnu (Amatucci et Sohl, 2004 ; Eddleston *et al.*, 2016 ; Nigam, Benetti et Mavoori, 2022 ; Snellman et Solal, 2023). D'autre part, à travers leur expérience de femme dirigeante fondatrice, les entrepreneures auront intégré les exigences du processus de financement, ce qui peut être apprécié par les investisseurs en capital. Chen, Yao et Kotha (2009) montrent que le degré d'engagement perçu (degré de préparation) influence positivement la décision d'investissement des capital-risqueurs. La passion et la ténacité affichées sont une bonne indication de l'engagement opérationnel des fondateurs d'entreprise, ce qui influence l'accès au financement (Harrison,

Mason et Smith, 2015 ; Maxwell et Lévesque, 2014 ; Murnieks, Cardon, Sudek, White et Brooks, 2016). Nous retenons ainsi que le niveau d'empouvoirement permet à l'entrepreneure d'afficher un degré d'engagement opérationnel, qui sera pris en compte par l'investisseur en capital. Il est donc possible de déduire que le parcours de dirigeante fondatrice entraîne une perception positive des capacités entrepreneuriales de la femme, ce qui va ensuite influencer favorablement la probabilité d'accéder à la levée de fonds.

Dans la littérature sur l'accès au prêt bancaire, peu d'études portent explicitement sur l'engagement opérationnel comme facteur de décision. Le conseiller bancaire, à travers son modèle, n'a pas la possibilité d'observer la mise en application des compétences opérationnelles des entrepreneures (Qi, Ongena et Cheng, 2022). Le secteur bancaire est particulièrement enclin à reproduire les stéréotypes de genre parmi ses conseillers, qui ne reconnaissent pas pleinement aux entrepreneures leur capacité à influencer les aspects opérationnels de leur entreprise (Marlow et Swail, 2014 ; Orhan, 2001). L'interprétation et les schémas de croyance des acteurs de la banque sont importants dans l'accès au crédit par les entrepreneurs (Bellucci, Borisov et Zazzaro, 2010). Le conseiller bancaire attend que les femmes se comportent conformément aux attentes sociales plutôt que de les encourager à démontrer leurs capacités en tant qu'entrepreneures à succès (Liu et Cowling, 2023). L'étude d'Eddleston *et al.* (2016) montre aussi une relation négative entre l'engagement (mesuré par un grand nombre d'heures consacrées à l'entreprise) et l'obtention du crédit bancaire. Les auteurs suggèrent en effet que, pour les banquiers, un grand nombre d'heures consacrées à l'entreprise peut vouloir dire que l'entrepreneur est inefficace ou mal adapté à la gestion de l'entreprise. Ainsi, l'empouvoirement des femmes, qui se traduit par la démonstration d'un engagement, risque d'accroître les préjugés à leur encontre dans le financement bancaire (Mascia et Rossi, 2017). Le parcours de la fondatrice dirigeante ne pourra pas jouer en sa faveur parce que, d'une part, le banquier ne peut pas observer son comportement futur et, d'autre part, le conseiller bancaire utilise un modèle social dominant qui désavantage l'entrepreneure.

Nous nous fondons alors sur les hypothèses suivantes :

**H1 :** le statut de dirigeante fondatrice influence la probabilité d'accès au financement par emprunt bancaire et par levée de fonds.

**H1a :** le statut de dirigeante fondatrice influence négativement la probabilité d'accès au financement par emprunt bancaire.

**H1b :** le statut de dirigeante fondatrice influence positivement la probabilité d'accès au financement par levée de fonds.

Dans le cadre du financement bancaire, les dirigeantes peuvent montrer leur engagement en utilisant, d'une part, les garanties financières et, d'autre part, l'implication du conjoint ou de la famille dans l'entreprise. Certaines entrepreneures ont témoigné qu'elles obtenaient facilement des prêts auprès de la banque grâce aux relations de leur mari avec le conseiller, ainsi qu'à la réputation favorable de leur famille (Chowdhury, Yeasmin et Ahmed, 2018). De plus, les hommes auraient une meilleure aptitude à négocier avec les conseillers bancaires (Mascia et Rossi, 2017), ce qui conduit les femmes à utiliser la présence du conjoint dans l'entreprise pour soutenir leur demande de financement. Selon Bellucci, Borisov et Zazzaro (2010), Madill,

Riding et Haines (2006) et Orhan (2001), les entrepreneures doivent apporter davantage de garanties dans leurs demandes de crédits bancaires. Basiglio et al. (2023) avancent que les entrepreneures sollicitent plus fréquemment un emprunt bancaire lorsque l'entreprise créée est la propriété de la famille, car cette dernière représente une garantie plus solide pour le banquier. Si la présence du conjoint influence positivement l'accès au financement bancaire, nous pouvons supposer qu'elle est associée à une faible confiance des dirigeantes fondatrices quant à leurs aptitudes cognitives et psychologiques. D'ailleurs, certaines recherches observent que la présence du conjoint dans l'entreprise peut être associée à un faible empouvoirement parce que le cercle familial ne constitue pas la meilleure ressource pour les défis entrepreneuriaux (Dong et Khan, 2023 ; Huis et al., 2020). Si l'investisseur a besoin de voir une dirigeante fondatrice engagée démontrant ses capacités (Chen, Yao et Kotha, 2009 ; Harrison, Mason et Smith, 2015), les recherches de Dong et Khan (2023) indiquent que la présence du conjoint influencera négativement la probabilité de lever des fonds.

La politique de rémunération peut également influencer la probabilité d'accès au financement. D'une part, le sacrifice que le dirigeant fondateur est capable de faire sur ses rémunérations personnelles afin d'investir peut constituer un élément favorable dans sa demande de financement auprès des investisseurs (Mitteness, Sudek et Cardon, 2012 ; Ola, Deffains-Crapsky et Dumoulin, 2019). D'autre part, Kerr et Nanda (2011) observent que la richesse personnelle est souvent associée à une plus grande capacité à surmonter les contraintes de financement et donc limite la propension à lever des fonds. Ces revenus peuvent agir comme un levier en permettant aux entrepreneurs d'investir suffisamment de capital dans leurs entreprises. Ils vont préférer un accès au crédit parce que leur revenu personnel leur sert de garantie auprès du banquier. Madill, Riding et Haines (2006) démontrent qu'un faible niveau de revenu et du niveau de garantie est significativement relié à un niveau élevé de refus de crédit bancaire aux femmes. Les difficultés d'accès au crédit semblent plus fortes dans des contextes sociaux où les femmes font l'objet de forte discrimination vis-à-vis du revenu par exemple (Malmström et al., 2023). Nous en déduisons donc les hypothèses suivantes :

**H2** : un revenu élevé influence la probabilité d'accès au financement par emprunt bancaire et par levée de fonds.

**H2a** : un revenu élevé influence positivement la probabilité d'accès au financement par emprunt bancaire.

**H2b** : un revenu élevé influence négativement la probabilité d'accès au financement par levée de fonds.

**H3** : la présence du conjoint dans l'entreprise influence la probabilité d'accès au financement par emprunt bancaire et par levée de fonds.

**H3a** : la présence du conjoint dans l'entreprise influence positivement la probabilité d'accès au financement par emprunt bancaire.

**H3b** : la présence du conjoint dans l'entreprise influence négativement la probabilité d'accès au financement par levée de fonds.

## 1.4. L'empouvoirement, le réseau professionnel et le financement externe

*1.4.1. L'empouvoirement des femmes entrepreneures et le réseau professionnel*

Le réseau constitue un soutien clé de développement des entreprises grâce à la cooptation sociale (Malmström *et al.*, 2020 ; Malmström et Wincent, 2018 ; Neumeyer, Santos, Caetano et Kalbfleisch, 2019). D'une part, la littérature met en lumière une difficile intégration des femmes aux réseaux d'affaires (Carrier, Julien et Menvielle, 2006 ; Barragan, Erogul et Essers, 2018). D'autre part, le contenu du réseau permet de distinguer les femmes qui sont en adéquation avec les valeurs dominantes de l'entrepreneuriat de celles qui manquent de légitimité dans l'environnement (McAdam, Harrison et Leitch, 2019). À travers le processus d'empouvoirement par l'expérience entrepreneuriale, les divers liens développés par certaines femmes leur permettent d'accéder aux ressources nécessaires au développement de leur entreprise (Constantinidis, 2010, 2021 ; Kelly et McAdam, 2023). Dong et Khan (2023) observent que chez les entrepreneures, le soutien social ne provient pas de la famille ni des proches, mais d'une multitude de relations permises par le nouveau statut de dirigeante. Elles obtiennent ainsi un fort soutien qui facilite l'accès aux ressources. D'ailleurs, en le faisant, elles acquièrent une capacité de persuasion et d'explication qu'elles remobilisent pour se faire coopter dans leur environnement d'affaires.

Historiquement, les entrepreneures sont plus susceptibles de nommer leur conjoint et leur famille en premier lorsqu'elles sont interrogées sur leur capital social (Powell et Eddleston, 2013). Elles y recherchent davantage un soutien moral et social plutôt qu'un accélérateur de *business*. Les réseaux exclusivement féminins exacerbent la ghettoïsation des femmes et l'effet des stéréotypes de genre alors que les réseaux mixtes permettent d'accumuler des ressources qui favorisent l'empouvoirement, l'amélioration de la capacité de choix et d'action en entrepreneuriat (McAdam, Harrison et Leitch, 2019). Neumeyer *et al.* (2019) montrent par ailleurs que les femmes entrepreneures ont un réseau d'entreprise de meilleure qualité que leurs homologues masculins. En outre, les femmes les plus expérimentées parviennent à établir un réseau connecté à plusieurs environnements sociaux diversifiés. La capacité à construire un réseau professionnel et mixte permet aux dirigeantes fondatrices de faire accepter socialement leur identité d'entrepreneure, capable de réussir (Barragan, Erogul et Essers, 2018 ; Kelly et McAdam, 2023).

### 1.4.2. Le réseau professionnel et les sources de financement

Le capital social constitue un indicateur important susceptible d'influencer l'accès au financement des entrepreneures, même si les résultats des études sont contradictoires. Carter et al. (2003) et Nigam, Benetti et Mavoori (2022) ne trouvent aucune influence du réseau sur l'utilisation des fonds propres et de la dette bancaire par les femmes entrepreneures. À l'inverse, Greene et al. (2001) affirment que les femmes sont impliquées dans des réseaux sociaux structurellement différents, ce qui peut entraîner un accès difficile aux capitaux propres par rapport aux entrepreneurs. Les entrepreneurs, qui sont en contact avec des professionnels expérimentés afin de signaler leur crédibilité, ont un accès facilité aux investisseurs (Murphy, Kickul, Barbosa et Titus, 2007). La même idée est défendue par Carter et al. (2003), qui avancent que le recours aux mentors amène les femmes à réduire l'utilisation des financements personnels et informels.

Les équipes mixtes ont un avantage dans l'accès au financement en raison de la qualité et de la confiance qu'elles inspirent, attirant ainsi un nombre plus important d'investisseurs individuels (Cicchiello et Kazemikhasragh, 2022). Des recherches montrent que les femmes vont développer une stratégie de compensation en intégrant des hommes issus de leur réseau dans leur équipe ou leur conseil afin de réduire les effets du stéréotype lié au genre (Cowling, Marlow et Liu, 2020 ; Qi, Ongena et Cheng, 2022). Brush et al. (2018) montrent que la mixité dans le comité exécutif d'une entreprise mature augmente les probabilités de lever des fonds auprès des capital-risqueurs. Les entrepreneures n'hésitent pas à évoquer les expériences des hommes présents dans leur comité de direction pour légitimer leur projet dans les levées de fonds (Alsos et Ljunggren, 2017 ; Murphy et al., 2007).

Peu d'études à notre connaissance expliquent le rôle du capital social sur le marché du crédit bancaire classique dans les pays développés. Les travaux sur le financement bancaire par les équipes de projet des entrepreneures semblent ignorer qu'une des caractéristiques de leur projet est l'implication de leur conjoint, proches et famille. Les conjoints sont cités comme premiers constituants du réseau des entrepreneures, même avant des experts conseillers (Orhan, 2001 ; Powell et Eddleston, 2013). En effet, la présence des proches dans l'équipe qui entoure une femme entrepreneure peut être un indicateur supplémentaire pris en compte par le conseiller bancaire. Par simple addition, l'engagement du conjoint et des membres de la famille va influencer la perception du risque. L'existence de proches, famille et amis peut renforcer ainsi sa légitimité (Powell et Eddleston, 2013). Cet effet n'est peu étudié dans la littérature en finance entrepreneuriale. Par ailleurs, le financement bancaire va aussi être favorisé par l'appartenance à des réseaux communautaires ou des réseaux strictement composés de femmes. Wellalage et Thrikawala (2021) affirment que les systèmes de crédit basés sur le relationnel au sein d'une communauté doivent permettre d'augmenter l'accessibilité au prêt bancaire. Perrin et Weill (2022) évoquent l'importance de la cooptation dans le cadre des modèles de microfinance, où la pression sociale des autres entrepreneures réduit le risque de défaut de remboursement. En effet, les femmes plus que les hommes sont plus sensibles aux coûts sociaux associés au non-remboursement d'un prêt (Lindvert, Patel et Wincent, 2017). Le secteur bancaire étant dominé par les stéréotypes de genre, les conseillers se représentent le réseau des femmes comme plus communautaire et familial. L'appartenance à ces groupes de paires n'a pas pour finalité première l'accès à des informations et compétences diversifiées, ce qui est plutôt le cas des réseaux professionnels. Par conséquent, les femmes avec des réseaux professionnels ne correspondent pas aux attentes sociales du secteur bancaire. Nous formulons donc les hypothèses suivantes :

**H4** : l'appartenance à un réseau professionnel influence la probabilité d'accès au financement par emprunt bancaire et par levée de fonds.

**H4a** : l'appartenance à un réseau professionnel influence négativement la probabilité d'accès au financement par emprunt bancaire.

**H4b** : l'appartenance à un réseau professionnel influence positivement la probabilité d'accès au financement par levée de fonds.

**H5** : l'appartenance à un réseau professionnel mixte influence la probabilité d'accès au financement par emprunt bancaire et par levée de fonds.

**H5a** : l'appartenance à un réseau professionnel mixte influence négativement la probabilité d'accès au financement par emprunt bancaire.

**H5b** : l'appartenance à un réseau professionnel mixte influence positivement la probabilité d'accès au financement par levée de fonds.

## 2. MÉTHODOLOGIE

### 2.1. Contexte de l'étude et modèle empirique

Cette recherche a été menée dans le cadre d'une enquête réalisée par Bpifrance Le Lab en 2022 auprès de 37 000 dirigeants et dirigeantes de petites ou moyennes entreprises (PME) et entreprises de taille intermédiaire (ETI) en France. Cette étude, qui totalise 1 871 réponses collectées auprès de 868 femmes et 1 003 hommes, vise à enrichir la compréhension des comportements des dirigeants au moyen d'une approche comparative. Pour notre analyse, nous avons décidé de nous concentrer sur l'accès au financement des dirigeantes d'entreprise, qui est un sujet peu exploré en France (Constantinidis, Cornet et Asandei, 2006 ; Liu et Cowling, 2023 ; Qi, Ongena et Cheng, 2022). L'accès au financement a été caractérisé à travers deux variables dépendantes principales : l'emprunt bancaire et la levée de fonds (Constantinidis, Cornet et Asandei, 2006 ; Aernoudt et De San José, 2020). Ce choix se veut représentatif des principales sources de financement externes explorées par les dirigeantes. Cette sélection est appuyée par une littérature académique soulignant l'importance de ces mécanismes dans la croissance et le développement des PME et ETI (Grandclaude, 2015). Après avoir sélectionné les données et défini nos variables dépendantes et de contrôle, nous avons identifié un total de 337 femmes entrepreneures dirigeantes fondatrices parmi un ensemble de 868 femmes répondantes. La répartition de cet échantillon est la suivante : 235 femmes dirigeantes fondatrices (70 %) ont été financées par emprunts bancaires et 102 (30 %) par levées de fonds.

Pour examiner la probabilité d'accès au financement en fonction des réponses au questionnaire, nous avons constitué une base de données comprenant principalement des variables binaires susceptibles d'influencer cet accès. Nous avons également utilisé des variables continues telles que l'âge, les revenus, le chiffre d'affaires, le nombre de réseaux, le nombre d'enfants, lorsque les données le permettaient. Initialement, quatre secteurs d'activité avaient été envisagés. Toutefois, en raison de la disponibilité limitée des données, seuls deux secteurs ont été retenus pour l'analyse. L'ensemble des données est regroupé dans le tableau 1.

Tableau 1. Présentation des variables dépendantes, indépendantes et de contrôle

| Caractéristiques des variables | Références au sein de la littérature | Variables | Mesure |
|---|---|---|---|
| Caractéristiques des variables dépendantes – type de financement | Constantinidis, Cornet et Asandei (2006) | Emprunt bancaire | Si l'entrepreneure est financée par un emprunt bancaire, la valeur prend 1, sinon 0 |
| | Constantinidis, Cornet et Asandei (2006) Aernoudt et De San José (2020) | Levée de fonds | Si l'entrepreneure est financée par une levée de fonds, la valeur prend 1, sinon 0 |
| Caractéristiques des variables indépendantes et de contrôle – femmes dirigeantes fondatrices | Mertzanis, Marashdeh et Ashraf (2023) | Dir. / Fon. | Si l'entrepreneure a le statut de dirigeante fondatrice, la valeur prend 1, sinon 0 |
| | Love, Nikolaev et Dhakal (2024) | Mariée | Si l'entrepreneure est mariée, la valeur prend 1, sinon 0 |
| | Constantinidis, Cornet et Asandei (2006) Aernoudt et De San José (2020) | Âge | L'âge de l'entrepreneure est représenté par la valeur du nombre d'années écoulées depuis sa naissance |
| | Love, Nikolaev et Dhakal (2024) | Enfants | Le nombre d'enfants de l'entrepreneure |

| Caractéristiques des variables | Références au sein de la littérature | Variables | Mesure |
|---|---|---|---|
| Caractéristiques des variables indépendantes et de contrôle – garantie dans la gestion de l'entreprise et garantie financière | Huis *et al.* (2020) | Conjoint | Si l'entrepreneure a son conjoint qui travaille dans l'entreprise qu'elle dirige, la valeur prend 1, sinon 0 |
| | Khaleque (2018) | Expérience | Si l'expérience de l'entrepreneure dans la gestion d'une entreprise est égale ou supérieure à dix années, la valeur prend 1, sinon 0 |
| | Constantinidis, Cornet et Asandei (2006) | Diplôme | Si l'entrepreneure a un niveau égal ou supérieur à bac +5, la valeur prend 1, sinon 0 |
| | Love, Nikolaev et Dhakal (2024) | Revenus | Si la rémunération annuelle de l'entrepreneure est égale ou supérieure à 50 k€, la valeur prend 1, sinon 0 |
| | Constantinidis, Cornet et Asandei (2006) | Taille Ent. | Si l'entrepreneure dirige une entreprise qui comprend un niveau égal ou supérieur à plus de 50 salariés, la valeur prend 1, sinon 0 |
| | Love, Nikolaev et Dhakal (2024) | CA Ent. | Si le chiffre d'affaires annuel de l'entreprise est égal ou supérieur à 10 millions d'euros, la valeur prend 1, sinon 0 |
| Caractéristiques des variables indépendantes et de contrôle – réseau, mixité du réseau et secteur d'activité | Constantinidis, Cornet et Asandei (2006) | Réseau | Le nombre de réseaux auxquels l'entrepreneure appartient est représenté par cette valeur |
| | (Neumeyer *et al.*, 2019) | Mixité | Si ce réseau professionnel est mixte, la valeur prend 1 sinon 0 |
| | Constantinidis, Cornet et Asandei (2006) | Secteur 1 | La variable Secteur 1 prend la valeur 1 si l'entreprise est dans l'industrie, le BTP ou le transport, sinon 0 |
| | Constantinidis, Cornet et Asandei (2006) | Secteur 2 | La variable Secteur 2 prend la valeur 1 si l'entreprise est dans les services, le commerce ou le tourisme, sinon 0. Ce secteur constitue le secteur de référence de notre analyse |

Nous avons mobilisé la régression logit, adaptée à l'analyse de variables dépendantes binaires (Hakmaoui et Yerrou, 2022), pour modéliser la probabilité d'accès au financement des entrepreneures. La régression logit présente l'avantage d'être flexible, capable de traiter des variables explicatives catégorielles (telles que le statut de dirigeante) et continues (telles que l'âge ou l'expérience de la dirigeante). En outre, elle permet d'évaluer au moyen d'effets marginaux l'impact d'une modification unitaire d'une variable indépendante sur la probabilité prédite de la variable dépendante, tout en maintenant constantes les autres variables du modèle.

Dans la formule du modèle de régression logistique pour l'accès au financement bancaire, chaque coefficient $\beta$ (bêta) est associé à une variable indépendante spécifique.

**Y** est la variable dépendante binaire (par exemple, l'accès au financement bancaire, 1 pour oui et 0 pour non).

$\beta_0$ est la constante (ou l'ordonnée à l'origine).

$\beta_1, \beta_2 \ldots \beta_k$ sont les coefficients des variables indépendantes $X_1, X_2, \ldots X_k$.

Ainsi, notre modèle pour l'accès à l'emprunt bancaire se définit de la manière suivante :

$$\log\left(\frac{P(Y_{1i} = 1)}{1 - P(Y_{1i} = 1)}\right) = \beta_0 + \beta_1 X_{1i} + \beta_2 X_{2i} + \cdots + \beta_k X_{ki} + \epsilon_i$$

Et notre modèle pour l'accès à la levée de fonds se définit de la manière suivante :

$$\log\left(\frac{P(Y_{2i} = 1)}{1 - P(Y_{2i} = 1)}\right) = \gamma_0 + \gamma_1 X_{1i} + \gamma_2 X_{2i} + \cdots + \gamma_k X_{ki} + \epsilon_i$$

## 2.2. Statistiques descriptives de l'échantillon et des variables

Pour introduire cette partie, nous mobilisons l'ensemble des informations qui nous apportent des précisions sur les coefficients de corrélation des variables dépendantes, indépendantes et de contrôle (Tableau 5 en annexe). La matrice de corrélation établie nous permet de confirmer l'absence de corrélation significative, au seuil de 5 %, entre la majorité des variables. Cette matrice montre également une faible corrélation négative (-0,27) entre les variables emprunts bancaires et levées de fonds, ce qui suggère que les entreprises qui contractent des emprunts bancaires ne sont pas les mêmes qui procèdent à des levées de fonds, et inversement.

Nous souhaitons par ailleurs présenter quelques données descriptives sur les entrepreneures dirigeantes fondatrices (Tableau 2). Notre échantillon total de 868 femmes entrepreneures est constitué de 337 dirigeantes fondatrices. L'âge moyen des entrepreneures dirigeantes fondatrices est d'approximativement 48 ans. Parmi elles, 53,32 % sont mariées. En ce qui concerne la maternité, 31,48 % d'entre elles ont plus d'un enfant, tandis que 13,27 % en ont un seul. L'accès au financement par emprunt bancaire concerne environ 69 % des dirigeantes fondatrices, tandis que 31 % ont accès à des levées de fonds. Les entrepreneures dirigeantes fondatrices mariées ayant un conjoint en entreprise sont légèrement plus enclines à choisir les emprunts bancaires. Ce choix pourrait être influencé par des considérations de stabilité ou de caution morale que le statut marital peut apporter lors des demandes de prêt. Les entrepreneures dirigeantes fondatrices plus jeunes et mieux diplômées (bac +5 et plus) tendent à se diriger vers les levées de fonds : 76 % des femmes ayant choisi cette option ont un niveau d'étude supérieur ou égal à bac +5, contre 44 % pour les emprunts bancaires. Un contraste est observé, où 56 % des dirigeantes fondatrices optant pour des emprunts bancaires ont de l'expérience, contre seulement 25 % pour les levées de fonds. Cela peut suggérer que les banques valorisent l'expérience antérieure pour la réduction du risque de crédit ou bien cette situation pourrait également indiquer que le financement par capitaux est privilégié par les femmes qui opèrent plus dans des secteurs technologiques et qui valorisent la croissance. Cette situation semble indiquer que celles qui optent pour les levées de fonds ont une meilleure capacité à mettre en œuvre des stratégies de croissance plus agressives, avec 60 % d'entre elles ayant des revenus plus élevés, comparativement à 43 % pour celles qui choisissent les emprunts bancaires. Par ailleurs, les entrepreneures qui lèvent des fonds sont généralement mieux connectées (au moins un réseau) et travaillent dans des environnements avec des réseaux davantage mixtes (43 %). Ces deux facteurs peuvent faciliter l'accès à des investisseurs diversifiés et à des opportunités de financement alternatif. Les entrepreneures dirigeantes fondatrices à la tête d'entreprises de plus petite taille ou réalisant un chiffre d'affaires moins élevé privilégient les levées de fonds, ce qui peut refléter une recherche de capital pour une croissance rapide ou pour surmonter des barrières financières initiales. Enfin, le secteur d'activité semble également influencer le type de financement.

TABLEAU 2. COMPARAISON DES CARACTÉRISTIQUES DES FEMMES ENTREPRENEURES SELON LE TYPE D'ACCÈS AU FINANCEMENT

|  | Emprunts bancaires | Levées de fonds |
|---|---|---|
| Mariée | 54 % | 40 % |
| Conjoint en entreprise | 32 % | 28 % |
| Âge | 48 ans | 44 ans |
| Nombre d'enfants | 1,077 | 1,125 |
| Égal ou supérieur à bac +5 | 44 % | 76 % |
| Expérience | 56 % | 25 % |
| Revenus | 43 % | 60 % |
| Nombre de réseaux | 0,8 | 1,18 |
| Mixité | 27 % | 43 % |
| Taille de l'entreprise | 41 % | 18 % |
| Chiffre d'affaires | 42 % | 17 % |
| Secteur d'activité 1 | 11 % | 28 % |
| Secteur d'activité 2 | 47 % | 68 % |
| Total | 235 femmes entrepreneures dirigeantes fondatrices | 102 femmes entrepreneures dirigeantes fondatrices |

## 3. RÉSULTATS : TYPOLOGIE DES FACTEURS QUI INFLUENCENT L'ACCÈS AU FINANCEMENT POUR LA DIRIGEANTE FONDATRICE

Les résultats et l'approche quantitative de cette étude jouent un rôle important dans la réponse à la problématique posée, en offrant une compréhension approfondie des facteurs influençant l'accès au financement pour les dirigeantes fondatrices.

Le premier facteur qui semble influencer l'accès au financement correspond au statut de dirigeante fondatrice. En effet, les résultats montrent que les femmes ayant le statut de dirigeante fondatrice ont une plus grande probabilité d'accéder à des levées de fonds tandis qu'il réduit les chances d'accéder à des emprunts bancaires (Tableau 3). Les modèles logistiques indiquent respectivement la présence d'un effet positif et significatif ainsi qu'un effet négatif et significatif (Tableau 3). Ce premier ensemble de résultats confirme la validité des hypothèses H1, H1a et H1b (Tableau 4).

Dans l'analyse initiale de l'engagement financier, nous observons dans un premier temps qu'un revenu élevé (égal ou supérieur à 50 k€) semble favoriser positivement l'accès à la levée de fonds avec un effet significatif qui invalide l'hypothèse H2b. Si la présence du conjoint au sein de l'entreprise semble influencer positivement l'accès à l'emprunt bancaire, ces résultats ne sont pas significatifs et ne peuvent confirmer l'hypothèse de Constantinidis et al. (2006) selon laquelle la présence d'un conjoint masculin dans l'entreprise peut accroître la confiance des conseillers bancaires. Par conséquent, les hypothèses H3, H3a et H3b sont réfutées.

Nos résultats positifs et significatifs indiquent que l'appartenance à un réseau professionnel accroît la probabilité d'obtenir des financements par emprunt bancaire, ce qui invalide l'hypothèse H4a. De la même manière, nous constatons également que l'appartenance à un réseau professionnel mixte accroît la probabilité d'obtenir des financements par levée de fonds, ce qui confirme l'hypothèse H5b. Cette observation met en lumière l'importance d'un tissu relationnel professionnel diversifié pour la mobilisation de ressources financières, corroborant les travaux de Kwapisz et Hechavarría (2018).

L'analyse de variables supplémentaires a révélé des indications notables, notamment le nombre d'enfants des dirigeantes fondatrices, qui montre une corrélation positive et significative avec l'accès à l'emprunt bancaire. En parallèle, une expérience égale ou supérieure à dix ans dans la gestion d'entreprise, ainsi qu'un chiffre d'affaires annuel égal ou supérieur à dix millions d'euros montrent un impact négatif et significatif en ce qui concerne la probabilité d'accès à une levée de fonds. Pour ce qui est de la probabilité d'accès à l'emprunt bancaire, seul le chiffre d'affaires révèle un impact négatif et significatif. Dans ce contexte, le CA élevé peut rassurer les bailleurs de fonds, car il constitue un indicateur de bonne santé financière, reflétant une capacité à générer des revenus substantiels et à assurer un niveau de compétitivité suffisant, réduisant le risque de probabilité d'obtenir un financement. Par ailleurs, nous pouvons constater que la détention d'un diplôme supérieur ou égal à bac +5 diminue la probabilité d'accès à l'emprunt bancaire. Dans le cas contraire, ce niveau de diplôme favorise l'accès à la levée de fonds. Les banques, lorsqu'elles évaluent les demandes de prêt, privilégient généralement la sécurité et la stabilité financière. Elles peuvent percevoir les titulaires de diplômes supérieurs comme ayant une propension plus élevée à prendre des risques. À l'inverse, les investisseurs en capital sont souvent à la recherche d'entreprises innovantes avec un fort potentiel de croissance, dans des secteurs comme la technologie ou la biotechnologie où un diplôme avancé est souvent un indicateur de capacité à innover et à développer de nouveaux produits ou services. Enfin, nous constatons un effet positif et significatif du secteur 1 pour l'accès à l'emprunt bancaire et à la levée de fonds.

En ce qui concerne l'analyse de la robustesse des résultats statistiques, le chi-2 teste l'indépendance entre les variables dans notre modèle. Les résultats significatifs (Prob. > chi-2 à 0,000 pour les deux modèles d'emprunt bancaire et de levée de fonds) suggèrent que le modèle est bien spécifié et que les variables incluses ont un lien fort avec la probabilité d'accès au financement. Le R² de McFadden offre une mesure de la qualité d'ajustement de notre modèle logistique. Avec des valeurs de R² à 0,3703 pour les emprunts bancaires et 0,3047 pour les levées de fonds, on observe une variabilité dans l'explicabilité des modèles. Le R² plus élevé pour les emprunts bancaires suggère que le modèle est mieux spécifié pour expliquer les facteurs influençant l'accès à ce type de financement par rapport à la levée de fonds. Cependant, le R² plus faible pour la levée de fonds indique que d'autres facteurs non inclus dans le modèle peuvent également jouer un rôle, suggérant des pistes pour des recherches futures.

Ces constatations offrent une vue d'ensemble sur l'examen de variables sélectionnées pour évaluer l'empouvoirement de femmes dirigeantes fondatrices de leur entreprise, mettant en lumière des aspects importants qui influencent leur accès au financement.

TABLEAU 3. PRÉSENTATION DES RÉSULTATS REPRÉSENTANT LA PROBABILITÉ D'ACCÈS AU FINANCEMENT POUR UNE FEMME ENTREPRENEURE DIRIGEANTE FONDATRICE

| Modèles | Logit | Effets marginaux | Logit | Effets marginaux |
|---|---|---|---|---|
| Variables | Emprunt bancaire | Emprunt bancaire | Levée de fonds | Levée de fonds |
| Dir. / Fon. | -3,100*** (-11,72) | -0,435*** (-16,91) | 2,151* (2,64) | 0,113* (2,62) |
| Mariée | 0,042 (0,22) | 0,005 (0,22) | -0,515 (-1,53) | -0,027 (-1,53) |
| Âge | 0,014 (1,23) | 0,001 (1,23) | -0,029 (-1,55) | -0,015 (-1,56) |
| Enfants | 0,524*** (4,23) | 0,073*** (4,38) | -0,100 (-0,61) | -0,005 (-0,61) |
| Conjoint | 0,295 (1,38) | 0,041 (1,38) | 0,258 (0,72) | 0,013 (0,72) |
| Expérience | 0,399 (1,93) | 0,056 (1,95) | -1,014* (-3,09) | -0,060* (-3,12) |
| Diplôme | -0,770*** (-4,08) | -0,108*** (-4,23) | 0,894* (2,64) | 0,047* (2,66) |
| Revenus | 0,204 (1,08) | 0,028 (1,08) | 0,997* (3,02) | 0,052* (3,05) |
| Taille Ent. | 0,109 (0,46) | 0,015 (0,46) | 0,738 (1,71) | 0,039 (1,70) |
| CA Ent. | -3,062*** (-10,92) | -0,430*** (-14,73) | -2,513*** (-5,60) | -0,133*** (-5,68) |
| Réseau | 0,325*** (3,40) | 0,045*** (3,48) | -0,117 (-0,93) | -0,006 (-0,93) |
| Mixte | 0,005 (0,03) | 0,005 (0,03) | 0,843* (2,66) | 0,044* (2,68) |
| Secteur 1 | 0,821** (2,28) | 0,115** (2,30) | 0,975* (2,65) | 0,051* (2,70) |
| Cons | 2,388*** (3,87) | - | -3,017* (-2,58) | - |
| R² | 0,3703 | - | 0,3047 | - |
| Log Likelihood | -378,34 | - | -158,84 | - |
| Prob. > chi-2 | 0,000*** | - | 0,000*** | - |
| OBS | 868 | - | 868 | - |

Notes : t-statistics entre parenthèses, *, **, *** ; seuils de significativité respectivement à 10 %, 5 % et 1 %.

TABLEAU 4. RÉCAPITULATIF DES PRINCIPAUX RÉSULTATS DE L'ACCÈS AU FINANCEMENT POUR UNE FEMME ENTREPRENEURE DIRIGEANTE FONDATRICE

| | Financement par levée de fonds | | | Financement par emprunt bancaire | | |
|---|---|---|---|---|---|---|
| | Signe de la relation | | Significativité | Signe de la relation | | Significativité |
| | Hypothèse | Résultat | | Hypothèse | Résultat | |
| Statut de femme dirigeante fondatrice | + | + | Oui (*) | - | - | Oui (***) |
| Statut de femme dirigeante fondatrice + revenu élevé | - | + | Oui (*) | + | + | Non |
| Statut de femme dirigeante fondatrice + conjoint dans l'entreprise | - | + | Non | + | + | Non |
| Statut de femme dirigeante fondatrice + appartenance réseau professionnel | + | - | Non | - | + | Oui (***) |
| Statut de femme dirigeante fondatrice + mixité du réseau professionnel | + | + | Oui (*) | - | + | Non |

## 4. DISCUSSION

Dans cet article, nous avons examiné empiriquement la manière dont l'empouvoirement personnel et relationnel des dirigeantes fondatrices de leur entreprise influence différemment les probabilités d'accès au financement par emprunt bancaire et par levée de fonds. Nos résultats montrent qu'il existe des caractéristiques qui augmentent ou limitent la probabilité d'accès à ces deux types de financement externe. Ces résultats contribuent à la littérature du financement entrepreneurial des femmes, trop souvent comparées aux hommes.

Notre recherche s'appuie sur la perspective selon laquelle l'empouvoirement constitue un point de départ des comportements entrepreneuriaux futurs (Digan *et al.*, 2019). Nous sommes partis de l'idée selon laquelle cet empouvoirement influence l'autoperception des compétences de l'entrepreneure, qui va influencer ses choix de financement. Son statut de dirigeante fondatrice renforce la crédibilité de son engagement opérationnel dans le projet, augmentant ainsi les chances de lever des fonds. Cette observation est conforme à la littérature sur les critères de décision d'investissement (Chen, Yao et Kotha, 2009 ; Harrison, Dibben et Mason, 1997 ; Mitteness, Sudek et Cardon, 2012 ; Ola, Deffains-Crapsky, 2019). Par ailleurs, la relation négative entre ce statut et l'accès aux emprunts bancaires est en accord avec les études qui soulignent la forte prégnance des stéréotypes de genre dans le secteur bancaire (Liu et Cowling, 2023 ; Mascia et Rossi, 2017).

L'engagement opérationnel est visible par le niveau de revenu de l'entrepreneure, qui influence positivement la probabilité d'accès à la levée de fonds. Les femmes qui veulent montrer un niveau d'empouvoirement suffisant doivent donc rester cohérentes dans leur stratégie en adoptant des standards de revenu existants dans l'environnement entrepreneurial. Si elles se sentent en mesure d'offrir des performances égales ou meilleures que celles des hommes, comme avancé par Aernoudt et De San José (2020) et Villaseca, Navío-Marco et Gimeno (2020), elles peuvent prétendre aux mêmes niveaux de revenu ; ceci peut même indiquer un bon niveau de confiance en soi. Il est essentiel que la politique de rémunération adoptée par ces entrepreneures renforce plutôt que ne diminue leur posture d'empouvoirement.

Par la suite, nous constatons que la présence du conjoint dans l'entreprise ne contribue pas à améliorer les chances d'obtenir un emprunt bancaire ou de lever des fonds. Cependant, le nombre d'enfants semble être un signal rassurant pour le banquier quant à la prudence au niveau de la gestion de l'entreprise et des prises de décision et, par ce fait, améliore la probabilité d'accès au financement bancaire (Chowdhury, Yeasmin et Ahmed, 2018). Si la présence du conjoint constitue un élément perturbateur dans le bon déroulement du projet de l'entrepreneure et limite son accès à la levée de fonds, nous constatons dans les résultats de notre étude que cet effet ne semble pas être avéré. Des recherches supplémentaires sur l'implication et le statut du conjoint en tant qu'associé pourraient être menées pour y vérifier d'éventuels effets négatifs. La littérature montre que la réussite du projet entrepreneurial et son accès à la levée de fonds est principalement déterminée par l'équipe et sa composition (Gompers, Gornall, Kaplan et Strebulaev, 2020).

La difficile intégration aux réseaux professionnels est une des préoccupations majeures de la littérature sur le financement des entrepreneures (Carrier, Julien et Menvielle, 2006 ; Malmström *et al.*, 2020 ; Malmström et Wincent, 2018). Nos résultats suggèrent toutefois que c'est la mixité au sein des réseaux, plutôt que leur simple appartenance, qui exerce une influence positive sur l'accès à la levée de fonds. Les investisseurs en capital valorisent la mixité des réseaux, perçue comme un indicateur de capacité à collaborer et à faire preuve d'ouverture d'esprit de la part des entrepreneures. Les femmes qui s'engagent dans des réseaux mixtes sont ainsi plus à même de constituer des équipes aux compétences variées, ce qui peut constituer un atout majeur (Brush *et al.*, 2018 ; Cicchiello et Kazemikhasragh, 2022 ; Cowling, Marlow et Liu, 2020 ; Gompers *et al.*, 2020). Notre étude va dans le sens des travaux d'Alsos et Ljunggren (2017) et Murphy *et al.* (2007) sur la légitimité sociale vis-à-vis du financement. Ainsi, un empouvoirement relationnel, accentué par la mixité, se révèle déterminant pour les levées de fonds (Anderson, John et Keltner, 2012 ; Thibaut et Kelley, 1959). Les réseaux non mixtes sont ceux où les entrepreneures vont rechercher de la sympathie, de l'affection ou du réconfort émotionnel (Malmström *et al.*, 2020 ; Malmström et Wincent, 2018). Dans le cas contraire, l'appartenance à des réseaux plutôt que sa mixité apparaît comme une stratégie nécessaire pour améliorer les chances d'obtenir un prêt bancaire. Les conseils et aides du réseau pour monter un projet ou un dossier de financement peuvent en effet faciliter les négociations avec les conseillers bancaires. Ainsi, nos résultats montrent un effet positif et significatif à l'appartenance à des réseaux professionnels sur la probabilité d'accès au financement bancaire, contrairement à l'hypothèse formulée. Cela vient donc nuancer la prégnance des stéréotypes de genre dans le secteur bancaire selon lesquels il n'est pas attendu que les femmes appartiennent à des réseaux professionnels.

Les résultats contrastés en fonction des deux types de financement externe nous permettent de participer au débat sur la négociation continue des femmes entre l'acceptation des normes masculines et la confrontation de ces dernières (Barragan, Erogul et Essers, 2018 ; Kelly et McAdam, 2023 ; Redien-Collot, Alexandre et Akouwerabou, 2022). La confrontation avec les normes masculines suppose que les entrepreneures s'affranchissent de la nécessaire présence du conjoint et mettent en avant d'autres caractéristiques, notamment leur appartenance aux réseaux pour accéder au financement bancaire.

Notre étude se situe dans cette nouvelle vague d'analyses empiriques, qui tente de comprendre l'unicité de la situation des entrepreneures, leur choix et leur expérience (Carrier, Julien et Menvielle, 2006 ; Ahl, 2006 ; Martinez Dy, 2021). Notre recherche contribue ainsi à la déconstruction du mythe autour du financement entrepreneurial des femmes. Le concept de l'empouvoirement que nous mobilisons nous amène à avancer que les entrepreneures peuvent être actrices de leur stratégie de financement. Elles peuvent prétendre à chacun des types de financement externe étudié. Leur expérience en tant que dirigeante fondatrice va les rendre plus crédibles vis-à-vis d'un financement externe. Ainsi, plutôt que de subir une discrimination ancrée dans les perceptions, certaines femmes font des choix qui vont contribuer à les rendre plus légitimes par rapport à d'autres, vis-à-vis des différents types de financement.

Les secteurs d'activité ne semblent pas discriminer les femmes dans l'accès à la levée de fonds selon nos résultats. Cela participe à nuancer des conclusions dans la littérature sur le manque d'ambition dans les projets de la part des entrepreneures (Cicchiello et Kazemikhasragh, 2022 ; Orhan, 2001). Concernant le financement par emprunt bancaire, les deux catégories de secteurs considérées ont des effets identiques, ce qui vient encore conforter la nuance que nous avons apportée.

En parallèle, notre recherche rejoint les conclusions de Flécher (2019), qui indiquent que les personnes ayant accès aux levées de fonds constituent une population très diplômée. Des formations spécifiques peuvent donc favoriser l'accès à des réseaux plus diversifiés et, par extension, à des opportunités de financement par levée de fonds. L'effet négatif de l'expérience sur la levée de fonds est cohérent avec la littérature, qui montre que ce sont plutôt les compétences démontrées par l'entrepreneure qui influencent la perception de l'investisseur (Maxwell et Lévesque, 2014 ; Murnieks et al., 2016). Nos résultats montrent ainsi que les investisseurs privilégient le diplôme alors que celui-ci a un effet négatif sur l'accès au financement bancaire. Le diplôme comme l'expérience peuvent être considérés comme des indicateurs indirects de l'engagement opérationnel des dirigeantes fondatrices. Il serait intéressant de mener des études plus approfondies sur l'interaction entre expérience et diplôme dans les démarches de financement des entrepreneurs.

## CONCLUSION

Dans le contexte de notre recherche, l'empouvoirement des femmes entrepreneures se manifeste par leur capacité à accéder à différentes sources de financement et par les décisions stratégiques qu'elles prennent en fonction de ces possibilités. Nos résultats révèlent les dynamiques d'empouvoirement sous-jacentes en illustrant comment les caractéristiques individuelles et professionnelles des femmes, telles que leur statut de dirigeante fondatrice, leur engagement opérationnel et financier, ainsi que leur réseau professionnel, impactent la probabilité d'accès au financement. En outre, notre recherche met en lumière l'importance de l'empouvoirement relationnel, à travers l'impact du réseau professionnel sur l'accès au financement.

Sur le plan théorique, notre recherche offre plusieurs contributions significatives. Premièrement, nous enrichissons la littérature sur l'entrepreneuriat des femmes en nous centrant sur l'expérience des femmes entrepreneures et en nous détachant des comparaisons habituelles avec les hommes (Nelson et Constantinidis, 2017 ; Martinez Dy, 2021). Si l'empouvoirement des femmes entrepreneures a fait l'objet d'une attention accrue, en particulier sur l'accès au microcrédit (Perrin et Weill, 2022), notre étude élargit ce cadre en s'intéressant à la manière dont l'empouvoirement influence l'accès à d'autres formes de ressources financières. Notre recherche révèle une probabilité que les choix de financement des femmes entrepreneures, ici les dirigeantes fondatrices, soient influencés par leurs relations professionnelles et réseaux, offrant ainsi une perspective plus nuancée sur la manière dont ces connexions affectent l'accès aux différentes sources de financement. En identifiant les mécanismes par lesquels l'empouvoirement relationnel et l'accès au financement sont liés, notre analyse offre des orientations pour les décideurs politiques et les praticiens qui cherchent à améliorer l'accès au financement pour les entrepreneures.

Notre étude, la première en France, offre une lecture enrichie de la disparité entre les entrepreneures. Elle met en relief la diversité des situations vécues, contribuant ainsi à déconstruire l'image homogène de l'entrepreneuriat des femmes très souvent dépeinte dans les recherches (Henry, 2021 ; Nelson et Constantinidis, 2017). Un apport significatif de notre étude réside dans son implication managériale pour les entrepreneures, notamment en mettant en lumière le type de financement le plus accessible selon leurs caractéristiques.

Enfin, si cette étude souligne la complexité des critères d'accès au financement et la nécessité d'une stratégie d'engagement financier et relationnel adaptée, elle pose des implications pour améliorer l'accès au financement pour les femmes entrepreneures, mettant en lumière le rôle des dispositifs et pratiques soutenant l'empouvoirement des femmes dans l'écosystème financier. Dans cet intérêt, il serait important de conduire une étude qualitative pour mieux comprendre les enjeux de l'empouvoirement relationnel dans la relation avec les financeurs.

ANNEXE 1. MATRICE DE CORRÉLATION

| Variables | Emprunts bancaires | Levées de fonds | Dir. / Fon. | Mariée | Âge | Enfants | Conjoint | Expérience | Diplôme | Revenus | Taille Ent. | CA Ent. | Réseau | Mixte | Secteur 1 | Secteur 2 |
|---|---|---|---|---|---|---|---|---|---|---|---|---|---|---|---|
| Emprunts bancaires | 1,0000 | | | | | | | | | | | | | | | |
| Levées de fonds | -0,2700* | 1,0000 | | | | | | | | | | | | | | |
| Dir. / Fon. | -0,2669* | 0,1679* | 1,0000 | | | | | | | | | | | | | |
| Mariée | 0,0417 | -0,0686* | -0,0009 | 1,0000 | | | | | | | | | | | | |
| Âge | 0,0185 | -0,1440* | -0,0614 | 0,1301* | 1,0000 | | | | | | | | | | | |
| Enfants | 0,1889* | 0,0728* | 0,1754* | 0,1344* | -0,1984* | 1,0000 | | | | | | | | | | |
| Conjoint | 0,0890* | 0,0001 | -0,0140 | 0,2477* | 0,0531 | -0,0387 | 1,0000 | | | | | | | | | |
| Expérience | 0,1022* | -0,1456* | -0,0519 | 0,1076* | 0,4301* | -0,0511 | 0,1232* | 1,0000 | | | | | | | | |
| Diplôme | -0,1725* | 0,1288* | 0,1333* | 0,0305 | -0,0904* | 0,0897* | -0,0734* | -0,1214* | 1,0000 | | | | | | | |
| Revenus | -0,0760* | 0,0755* | 0,1407* | -0,0181 | 0,0114 | -0,0077 | 0,0354 | -0,1053* | -0,0333 | 1,0000 | | | | | | |
| Taille Ent. | -0,0024 | -0,1332* | -0,4894* | 0,0346 | 0,0544 | -0,1582* | 0,0280 | 0,0913* | -0,0767* | -0,1527* | 1,0000 | | | | | |
| CA Ent. | -0,4222* | -0,2756* | -0,3402* | -0,0018 | 0,1164* | -0,2848* | -0,0443 | 0,0073 | -0,0007 | 0,1031* | 0,4047* | 1,0000 | | | | |
| Réseau | 0,2700* | 0,1634* | 0,2009* | 0,0046 | -0,0405 | 0,1878* | -0,0099 | -0,0050 | 0,0723* | 0,0108 | -0,1521* | -0,4414* | 1,0000 | | | |
| Mixte | 0,0869* | 0,1357* | -0,0474 | -0,0388 | -0,0271 | 0,0239 | 0,0057 | -0,1116 | 0,1080* | 0,0186 | 0,1083* | -0,0909* | 0,1954* | 1,0000 | | |
| Secteur 1 | 0,1466* | 0,2157* | 0,1730* | 0,0056 | -0,0072 | 0,0507 | 0,0304 | -0,0176 | 0,0169 | 0,0702* | -0,1327* | -0,2601* | 0,2363* | 0,1054* | 1,0000 | |
| Secteur 2 | 0,3111* | 0,2185* | 0,3691* | 0,0256 | -0,0798* | 0,3094* | 0,0478 | -0,0260 | 0,0162 | 0,0287 | -0,3916* | -0,5877* | 0,3525* | 0,0119 | -0,2006* | 1,0000 |

\* : significatif respectivement au seuil de 5 %.

# RÉFÉRENCES